\documentclass[twocolumn,prb,aps,floatfix,superscriptaddress,amssymb]{revtex4}
\usepackage{graphicx}
\usepackage{amsmath,amssymb,amsfonts}
\usepackage{color}
\usepackage{dsfont}
\usepackage{bm}
\usepackage[normalem]{ulem}
\usepackage{hyperref}
\hypersetup{
    colorlinks,%
    citecolor=blue,%
    filecolor=blue,%
    linkcolor=blue,%
    urlcolor=blue
}

\begin{document}  
\title{Electron-electron interaction mediated indirect
  coupling of electron and magnetic ion or nuclear spins in self-assembled quantum dots}

\author{Udson C. Mendes} 
\affiliation{Quantum Theory Group, Security and Disruptive
  Technologies, National Research Council, Ottawa, Canada K1A0R6}   
\affiliation{Institute of Physics ``Gleb Wataghin'', State University
  of Campinas, Campinas, S\~ao Paulo, Brazil} 

\author{Marek Korkusinski} 
\affiliation{Quantum Theory Group, Security and Disruptive
  Technologies, National Research Council, Ottawa, Canada K1A0R6}
 
\author{Pawel Hawrylak}  
\affiliation{Quantum Theory Group, Security and Disruptive
  Technologies, National Research Council, Ottawa, Canada K1A0R6} 
\affiliation{Department of Physics, University of Ottawa, Ottawa,
  Canada}
  \affiliation{WPI-AIMR, Tohoku University, Sendai,
  Japan} 
     
\begin{abstract} 
We show here the existence of the indirect coupling of electron and
magnetic or nuclear ion spins in self-assembled quantum dots mediated by
electron-electron interactions.  
With a single localized spin placed in the center of the dot, only the
spins of electrons occupying the zero angular momentum states couple
directly to the localized spin. 
We show that when the electron-electron interactions are included,
the electrons occupying finite angular momentum orbitals interact with 
the localized spin. 
This effective interaction is obtained using exact diagonalization of
the microscopic Hamiltonian as a function of the number of electronic
shells, shell spacing, and anisotropy of the electron-Mn exchange interaction. 
The effective interaction can be engineered to be either ferromagnetic
or antiferromagnetic by tuning the parameters of the quantum dot. 
\end{abstract}

\maketitle 

\section{Introduction}
There is currently interest in understanding the coupling of a
localized spin, either magnetic impurity or nuclear spin, with spins of interacting
electrons.\cite{gaj_kossut_book2011,koenraad_flatte_nat2011} This includes the Kondo effect in metals
\cite{kondo_ptp1964,hewson_book,madhavan_wingreen_science1998,li_delley_prl1998,jacob_palacios_prb2013} 
and quantum dots,\cite{gordon_kastner_nat1998,cronenwett_kouwenhoven_science1998,
kleemans_koenraad_natphys2010,latta_imamoglu_nat2011} 
the impurity spin in diamond,\cite{balasubramanian_chan_nature2008,mamin_kim_science2013} 
charged quantum dots with magnetic ions,
\cite{govorov_prb2004,qu_hawrylak_prl2005,leger_besombes_prl2006,nguyen_peeters_prb2008,mendes_korkusinski_prb2013} 
and nuclear spins coupled to fractional quantum Hall states.
\cite{machida_hirayama_prb2002,yusa_hirayama_nature2005,fauzi_hirayama_apl2012,akiba_hirayama_prb2013}
Here we focus on a highly tunable system of quantum dots with a single magnetic
ion and a controlled number of electrons. Such a system is realized in CdTe quantum dots
 with a magnetic impurity
in the center of the dot loaded with a controlled, small at present, number of
electrons.\cite{leger_besombes_prl2006} 
The interplay between electron-electron Coulomb interactions
and the electron-Mn exchange interaction has been studied using exact
diagonalization techniques\cite{qu_hawrylak_prl2005,nguyen_peeters_prb2008} 
and using the mean-field
approach.\cite{govorov_prb2004,govorov_kalameitsev_prb2005} 
Other studies focused on electron-electron interactions in excitonic
complexes coupled with localized 
spins.\cite{leger_besombes_prl2006,rossier_prb2006,govorov_kalameitsev_prb2005,
trojnar_korkusinski_prl2011,trojnar_korkusinski_prb2013,mendes_korkusinski_prb2013}

Here we focus on the indirect coupling of electron and magnetic or nuclear ion spins 
in self-assembled quantum dots (QDs) mediated by the electron-electron interaction.
With a localized spin placed in the center of the dot, only the spins of
electrons occupying the zero angular momentum states of the $s$, $d$,
$\ldots$ shells couple directly to the localized spin via a contact
exchange interaction. 
The situation is identical to the Kondo problem in metals where only
zero angular momentum states of the Fermi sea are considered  as
interacting with the localized spin. 
The question arises as to the role of electron-electron interactions. 
Here we show that, in quantum dots, when electron-electron
interactions are included, the electrons occupying finite angular
momentum orbitals (e.g., $p$ shell) do interact with the localized spin. 
The effective interaction for $p$-shell electrons is obtained using
exact diagonalization of the microscopic Hamiltonian as a function of
the number of electronic shells, shell spacing and anisotropy of
the exchange interaction. 
The anisotropy of exchange interpolates between the interaction types
characteristic for conduction band electrons (Heisenberg-like) and
valence band holes (Ising-like). We show that the effective 
electron-electron mediated exchange interaction can be engineered 
to be either ferro- or antiferromagnetic by varying quantum dot parameters.  

The paper is organized as follows: In Sec. II we describe the model
of a self-assembled quantum dot with a single Mn impurity in its center
and a controlled number of electrons. 
Section III presents results of exact diagonalization of the model
Hamiltonian for quantum  dots confining from two to six electrons and
the emergence of the indirect electron-Mn coupling for QDs with a 
partially filled $p$ shell. Section IV summarizes our results.  

\section{Model}

\begin{figure}
\begin{center}
\includegraphics[width=0.48\textwidth]{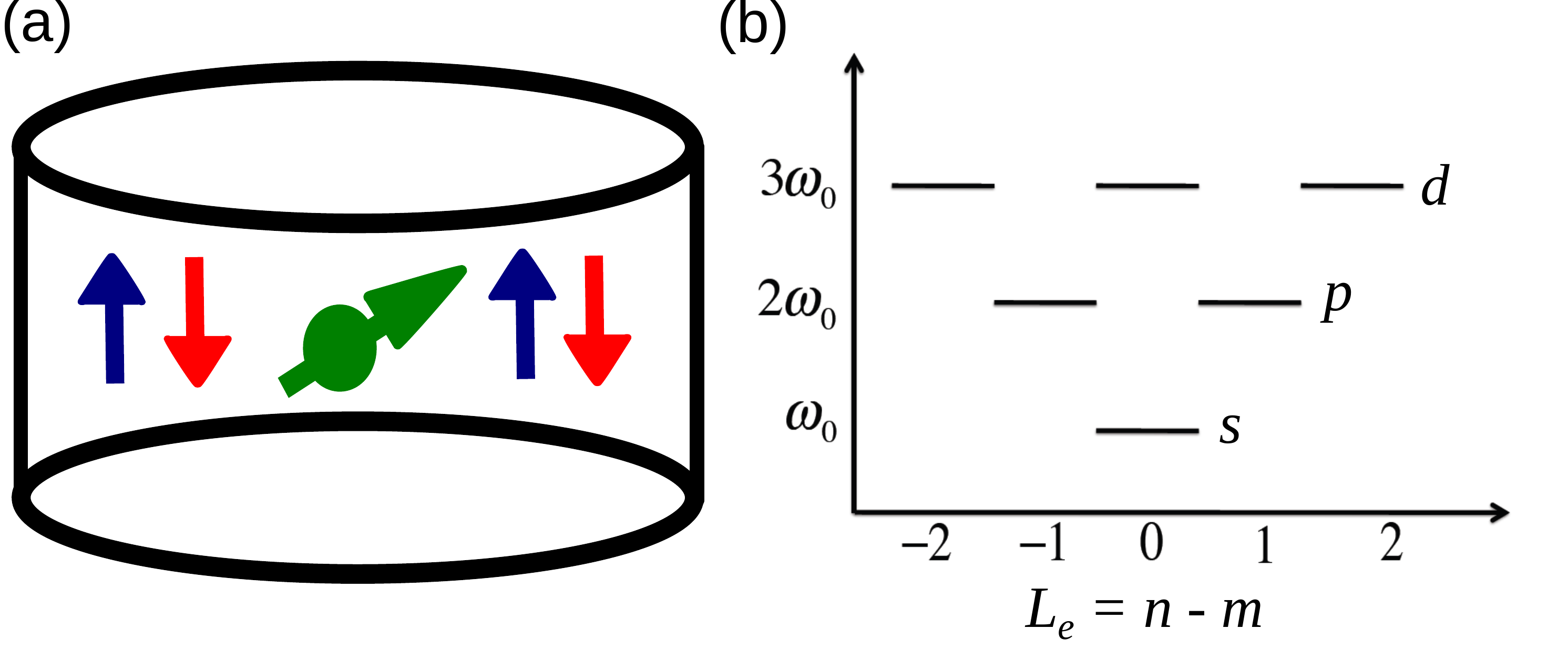}
\caption{(Color online) (a) Schematic representation of a CdTe 
quantum dot containing electrons and one Mn spin at its center. 
  (b) Single-particle states as a function of angular momentum.  
  \label{fig0}}
\end{center}
\end{figure}

We consider a model system of $N$ electrons ($N=2,\ldots,6$) confined
in a two-dimensional (2D) parabolic quantum dot with a single magnetic 
impurity in the center. Figure \ref{fig0}(a) illustrates a 
schematic representation of the investigated QD. For definiteness we 
consider an isoelectronic impurity, a manganese ion with a total spin $M=5/2$ 
in a CdTe quantum dot.\cite{gaj_kossut_book2011} In the effective mass and 
envelope function approximations, the single-particle states $|i,\sigma \rangle$ 
are those of a 2D harmonic oscillator (HO) with the characteristic frequency 
$\omega_0$. They are labeled by two orbital quantum numbers, $i=\{n,m\}$, 
and the electron spin $\sigma=\pm1/2$. The single-particle states are 
characterized by energy $E_{n,m}=\omega_{0}(n+m+1)$ and angular momentum 
$L_{e}=n-m$. Figure \ref{fig0}(b) shows the single-particle 
states as a function of angular momentum. We express all energies in 
units of the effective Rydberg, Ry$^{*} = m^{*} e^{4}/2\epsilon^{2} \hbar^{2}$, 
and all distances in units of the effective Bohr radius, 
$a_{B}^{*}=\epsilon \hbar^{2} /m^{*}e^{4}$, where $m^*$, $e$, $\epsilon$,  
and $\hbar$ are respectively the electron effective mass and charge, the 
dielectric constant, and the reduced Planck constant. For CdTe we take 
$m^{*}=0.1m_{0}$ and $\epsilon=10.6$, where $m_{0}$ is the free-electron mass, 
and Ry$^{*} =12.11$ meV and $a_{B}^{*}=5.61$ nm. Unless otherwise stated, 
we take the HO frequency $\omega_{0}=1.98$ Ry$^{*}$, consistent with our 
previous work.\cite{trojnar_korkusinski_prb2013} 
%%%%%%%%%%%%%%%%%%%%Figure 0 %%% QD single-partice %%%%%%%%%%%%%%%%
%%%%%%%%%%%%%%%%%%%%%%%%%%%%%%%%%%%%%%%%%%%%%%%%%%%%%%%%

The Hamiltonian of $N$ electrons confined in our QD and interacting
with a single Mn spin is written as \cite{qu_hawrylak_prl2005}
%%%%%%%%%%%%%%%%%%%%%%%%%%%%%%%%%%%%%%%%%%%%%%%%%%%%%%%%%%%%%%%%%%%%%%%%%%%
\begin{align} \label{eq1}
H &=  \sum_{i,\sigma} E_{i,\sigma} c_{i,\sigma}^{\dagger} c_{i,\sigma} + \frac{\gamma}{2} \sum_{\substack{i,j,k,l \\ \sigma, \sigma^{\prime} }} 
\langle i,j |V_{ee}| k,l \rangle c_{i,\sigma}^{\dagger} c_{j,\sigma^{\prime} }^{\dagger}c_{k,\sigma^{\prime} }c_{l,\sigma} \nonumber \\ 
&-\sum_{i,j} \frac{J_{i,j}(R)}{2}\left[ \left(c_{i,\uparrow}^{\dagger} c_{j,\uparrow} - c_{i,\downarrow}^{\dagger} c_{j,\downarrow}
\right)M_{z} + \varepsilon \left( c_{i,\downarrow}^{\dagger} c_{j,\uparrow}M^{+}  \right. \right.  \nonumber \\
& \left.  \left. + c_{i,\uparrow}^{\dagger} c_{j,\downarrow}M^{-} \right) \right],
\end{align}
%%%%%%%%%%%%%%%%%%%%%%%%%%%%%%%%%%%%%%%%%%%%%%%%%%%%%%%%%%%%%%%%%%%%%%%%%%
where $c_{i,\sigma}^{\dagger}$ ($c_{i,\sigma}$) creates (annihilates)
an electron on the orbital $i=\{m,n\}$ with spin $\sigma$.  

In the above Hamiltonian, the first term is the single-particle
energy and the second term is the electron-electron ($e$-$e$) Coulomb
interaction. 
The $e$-$e$ term is scaled by a dimensionless parameter $\gamma$:
$\gamma=0$ describes the noninteracting electronic system 
and $\gamma=1$ describes the interacting system.
The matrix elements $\langle i,j |V_{ee}| k,l \rangle$ of the Coulomb
interaction are evaluated in the basis of 2D HO orbitals in the closed
form.\cite{hawrylak_ssc1993}

The last term of the Hamiltonian describes the electron-Mn  
interaction ($e$-Mn). 
It is scaled by the exchange coupling matrix elements $J_{i,j}(R) =
J_{C}^{\text{2D}} \phi_{i}^{*}(R)\phi_{j}(R)$, where
$J_{C}^{\text{2D}} = 2J_{\text{bulk}}/d$,
$J_{\text{bulk}} = 15$ meV nm$^{3}$ is the \textit{s-d} exchange  
constant for the CdTe bulk material, $d=2$ nm is the QD height, 
and $\phi_{i} (R)$ is the amplitude of the HO wave function at  
the Mn position $R$. In particular, we define 
$J_{ss}(R) = J_{C}^{\text{2D}} \phi_{s}^{*}(R)\phi_{s}(R)$, which is 
the matrix element of an electron on the $s$ shell interacting 
with a magnetic ion. For Mn at the QD center its value is $J_{ss} 
\approx 0.15$ meV. 

The $e$-Mn interaction consists of two terms. The first one is the 
Ising interaction between the electron and Mn spin.The second term accounts 
for the $e$-Mn spin-flip interactions. The anisotropy of the exchange 
interaction is tuned by the factor $\varepsilon$. By setting $\varepsilon=0$ 
we obtain the anisotropic Ising $e$-Mn exchange Hamiltonian and setting 
$\varepsilon=1$ we obtain the isotropic, Heisenberg exchange Hamiltonian. 
In the former case, the spin projections $s_{z}$ and $M_{z}$ are separately 
good quantum numbers. The total spin projection of the electrons depends 
on the number and polarization of the particles. For the manganese spin 
we have $M=5/2$ and the six possible spin projections $M_z=-5/2,\ldots,5/2$. 
The isotropic Heisenberg Hamiltonian, in contrast, conserves the total 
angular momentum $\mathbf{J}=\mathbf{M}+\mathbf{S}$ and its projection
$J_{z}=s_{z}+M_{z}$. Hence, for the case $\varepsilon=1$, one can 
establish the total spin quantum number $J$ of the given manifold 
of states by considering its degeneracy $g(J)=2J+1$.

Since the elements $J_{i,j}$ depend on the position $R$ of the
Mn spin, the $e$-Mn coupling can be engineered
by choosing a specific $R$.\cite{qu_hawrylak_prl2005}
In this work we place the Mn spin in the center of the QD and the only
nonzero matrix elements $J_{i,j}$ appear if both orbitals $i$ and $j$
are zero angular momentum states.
The spin of an electron placed on any other HO orbital is not coupled
directly to the Mn spin.

The eigenenergies and eigenstates of the Hamiltonian (\ref{eq1}) are obtained
in the configuration-interaction approach.
In this approach, we construct the Hamiltonian matrix in the basis of
configurations of $N$ electrons and one Mn spin:
$|\nu_{i} \rangle = |i_{1\uparrow}, i_{2\uparrow},\ldots
,i_{N\uparrow}\rangle|j_{1\downarrow}, j_{2\downarrow},\ldots ,  
j_{N\downarrow}\rangle|M_{z}\rangle$, 
where $|i_{1\sigma},
i_{2\sigma},\ldots , i_{N\sigma}\rangle = c_{i_{1\sigma}}^{\dagger} 
c_{i_{2\sigma}}^{\dagger}, \ldots, c_{i_{N\sigma}}^{\dagger}
|0\rangle$, $|0\rangle$ is the vacuum state, and $N = N_{\uparrow} +
N_{\downarrow}$  
is the number of electrons, in which $N_{\uparrow}$ and
$N_{\downarrow}$ are the number of electrons with spin up and spin
down, respectively. 
The total number of configurations depends on the number of electrons
and on the number of the HO shells available in the QD.
With Mn impurity in the center, the total orbital
angular momentum of electrons $L = \sum_{i=1}^{N} L_{e}^{i}$ is
conserved by the Hamiltonian (\ref{eq1}).
Moreover, depending on the anisotropy of $e$-Mn interactions, the
Hamiltonian also conserves the total projections $S_z$ and $M_z$ of
the electron and Mn spin separately (the Ising model) or the
projection $J_{z}=s_{z}+M_{z}$ of the total spin (the Heisenberg
model). 
Based on these conservation rules, we divide the basis of
configurations into subspaces labeled by the numbers $L$, $S_z$, and
$M_z$ (for the Ising model) or $L$ and $J_z$ (for the Heisenberg
model), and diagonalize the Hamiltonian in each subspace separately.

Our model is also suitable for electrons interacting 
with a single nuclear spin. In the Fermi-contact hyperfine interaction,
\cite{urbaszek_imamoglu_rmp2013} the Hamiltonian of electrons interacting 
with nuclear spins has the same form as the Hamiltonian of electrons 
interacting with Mn spins. Even though the interaction between electrons 
and single nuclear spins has not been achieved in self-assembled quantum 
dots, today it is possible to manipulate a few nuclear spins in 
diamond,\cite{lee_wrachtrup_natnanotech2013} silicon,
\cite{morello_thewalt_nature2010,saeedi_thewalt_science2013} and 
carbon nanotubes.\cite{churchill_marcus_natphys2013}

The computational procedure adopted in this work is as follows.
For a chosen number of electrons $N=2,\ldots,6$ and a chosen number
of HO shells, we look for the ground and several excited states for
the system with and without $e$-$e$ interactions in the Ising and
isotropic Heisenberg models.
By analyzing the degeneracies of the states we find the total spin of
the system.
Further, from the ordering of different states with respect to their
total spin we draw conclusions as to the ferromagnetic or
antiferromagnetic character of the effective $e$-Mn interactions.
By comparing the results for the system with and without the $e$-$e$
interactions ($\gamma=1$ or $\gamma=0$, respectively) we 
establish the $e$-$e$ interaction mediated effective $e$-Mn Hamiltonian
for electrons not directly coupled to the central spin.

\section{Spin singlet closed shells coupled with the magnetic ion}

We start with a discussion of a filled $s$ shell with $N=2$
electrons in the zero angular momentum channel.  
Each electron is directly coupled to the Mn impurity, but the singlet
state couples only via $e$-$e$ interactions.\cite{trojnar_korkusinski_prb2013}
Here we discuss the role of anisotropy of the exchange interaction on this
indirect coupling. A similar discussion applies to other closed shells, 
e.g., $N=6$.

\begin{figure}
\begin{center}
\includegraphics[width=0.48\textwidth]{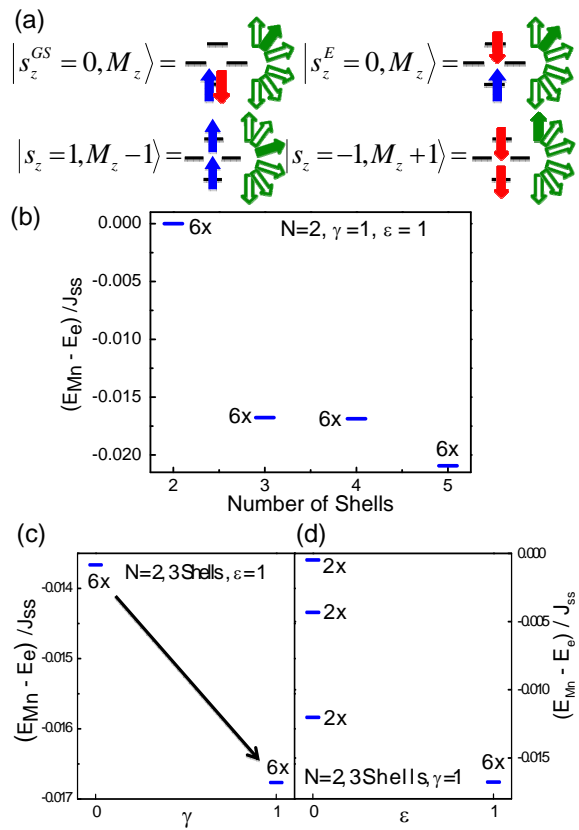}
\caption{(Color online) (a) Schematic pictures of two-electron-Mn
  configurations, GS and electronic triplet states, coupled by the
  $e$-Mn interactions.
  (b) Ground-state energy of the two-electron-Mn system 
  as a function of the number of quantum-dot shells. 
  (c) and (d) Ground-state energies of the two-electron-Mn system for
  the quantum dot confining three shells plotted as a function of the
  strength of electron-electron interactions in the Heisenberg $e$-Mn
  model (c) and as a function of the isotropy of the $e$-Mn Hamiltonian
  for the fully interacting electron system (d).
  Numbers at the energy level bars represent the degeneracy of
  states. 
  \label{fig1}}
\end{center}
\end{figure}

The lowest-energy $s$-shell spin singlet configuration with $S=0$ and
orbital angular momentum $L=0$,
$|s_{z}^{\text{GS}}=0,M_{z}\rangle=c_{s,\uparrow}^{\dagger}c_{s,\downarrow}^{\dagger}|0,M_{z}\rangle$, 
is shown schematically in the top left panel of Fig.~\ref{fig1}(a).
The expectation value of the $e$-Mn Hamiltonian against the
configuration $|s_{z}^{\text{GS}}=0,M_{z}\rangle$ is zero.
%%%%%%%%%%%%%%%%%%%%Figure 1 %%% s=shell %%%%%%%%%%%%%%%%
%%%%%%%%%%%%%%%%%%%%%%%%%%%%%%%%%%%%%%%%%%%%%%%%%%%%%%%%

Increasing the number of confined shells to three adds one additional 
orbital $(1,1)$ with zero angular momentum in the $d$-shell directly
coupled to the Mn spin.
Now the two-electron triplet states with total angular momentum $L=0$
couple to the Mn spin.
The triplet with $S_z=0$, $|s_z^{E}=0,M_z\rangle 
= (1/\sqrt{2}) \left(c_{d,\uparrow}^{\dagger}c_{s,\downarrow}^{\dagger}
- c_{s,\uparrow}^{\dagger} c_{d,\downarrow}^{\dagger} \right)|0\rangle
|M_z\rangle$. One of its components is shown schematically in the top right 
panel of Fig.~\ref{fig1}(a), while the bottom left panel of that figure shows
the spin-polarized triplet 
$|s_{z}=1,M_{z}-1\rangle
=c_{s,\uparrow}^{\dagger}c_{d,\uparrow}^{\dagger}|0,M_{z}-1\rangle$,
and the bottom right panel shows the triplet
$|s_{z}=-1,M_{z}+\rangle
=c_{s,\downarrow}^{\dagger}c_{d,\downarrow}^{\dagger}|0,M_{z}+1\rangle$.
Applying the $e$-Mn Hamiltonian to the $|s_{z}^{\text{GS}}=0,M_{z}\rangle$ state, we
obtain 
%%%%%%%%%%%%%%%%%%%%%%%%%%%%%%%%%%%%%%%%%%%%%%%%
\begin{align} \label{eq2}
&H_{e\text{-}Mn}|s_{z}^{GS}=0,M_{z}\rangle = 
-\frac{J_{sd}}{\sqrt{2}}M_{z}|s_z^{E},M_{z}\rangle \nonumber \\
&- \frac{J_{sd}}{2} \varepsilon 
\left(\beta_{-} |s_{z}=1,M_{z}-1\rangle 
- \beta_{+} |s_{z}=-1,M_{z}+1\rangle \right),
\end{align}
%%%%%%%%%%%%%%%%%%%%%%%%%%%%%%%%%%%%%
where $J_{sd}$ is the exchange matrix element in which one electron is
scattered from the $s$ orbital to the $d$ orbital and 
$\beta_{\pm} = \sqrt{(M\mp M_{z})(M\pm M_{z} +1)}$. 
We find that upon the inclusion of the $d$ shell, the low-energy
$s$-shell singlet two-electron configuration becomes coupled by $e$-Mn
interactions to electron triplet configurations, with and without
flip of the Mn spin.

We now diagonalize the two-electron-Mn Hamiltonian and compute the
ground-state (GS) energy $E_{\text{Mn}}$ of the QD with a manganese ion, and 
the energy $E_e$ of the system without Mn.
Figure~\ref{fig1}(b) shows the effect of the Mn ion on the ground state
energy, $\Delta = (E_{\text{Mn}}-E_e)/J_{ss}$, measured from the ground-state
energy without the Mn ion, as a function of the number of shells for the 
interacting system ($\gamma=1$) and the isotropic exchange interaction
($\varepsilon=1$).  
We find that, irrespective of the number of confined shells, the GS is
sixfold degenerate, with the total spin $J_{\text{GS}}=5/2$.
However, the energy of the GS markedly depends on the number of 
shells.
For two confined shells we have $\Delta=0$, because in this case 
we can generate only one configuration, $|s_z^{\text{GS}}=0,M_z\rangle$,
which is decoupled from the Mn spin.
The inclusion of the $d$ shell  adds an additional $L_e=0$
orbital into the single-particle basis, resulting in the scattering of
electrons by the localized spin and lowering of energy.
A further lowering of the energy occurs when the fifth shell,
containing another $L_e=0$ single-particle state, becomes confined.

Now we fix the number of shells to three, set the Heisenberg form of
$e$-Mn interactions and study the effect of $e$-$e$ interactions.
Figure~\ref{fig1}(c) shows the energy $\Delta$  without
($\gamma=0$) and with full Coulomb interactions ($\gamma=1$).
We find that the ground state in both cases is sixfold degenerate but
the $e$-$e$ Coulomb interactions enhance the effects of the $e$-Mn
coupling, lowering $\Delta$. 
This is due to a larger contribution of triplet configurations to the
GS. 

We now compare the results for the isotropic coupling versus the anisotropic 
coupling. For the anisotropic coupling, $\varepsilon=0$, we observe that the 
GS is split into three energy levels labeled by $|M_{z}|$, each of them twice 
degenerate, as shown in Fig.~\ref{fig1}(d). In Ising-like coupling the total 
angular momentum $J$ is not conserved, and the characteristic sixfold degeneracy 
of the ground state is broken. Comparing the isotropic and anisotropic coupling 
we observe that $\Delta$ is negative for both couplings and also that the 
Heisenberg-like interaction results in a lower energy than the Ising-like 
interaction.\cite{trojnar_korkusinski_prb2013}

\section{Electrons in finite angular momentum channels}

In this section we discuss electrons populating finite angular
momentum channels which are not directly coupled with the Mn ion. 
For $N=3$ we show the existence of an effective coupling mediated by
$e$-$e$ interactions. Similar results are obtained for $N=5$.  

%%%%%%%%%%%%%%%%%%%%%%%%%%%%%%%%%%%%%%%%%%%%%%%%%%%%%%%%%%%
\subsection{ One electron on the $p$ shell}

The lowest-energy configuration in the ground state of three electrons
is formed by two electrons in the $s$ shell and one electron in the
$p$ shell. With Mn in the QD center the total angular momentum $L$ of 
the three electrons is conserved and we show the results for $L=1$.  

\begin{figure}
\begin{center}
\includegraphics[width=0.48\textwidth]{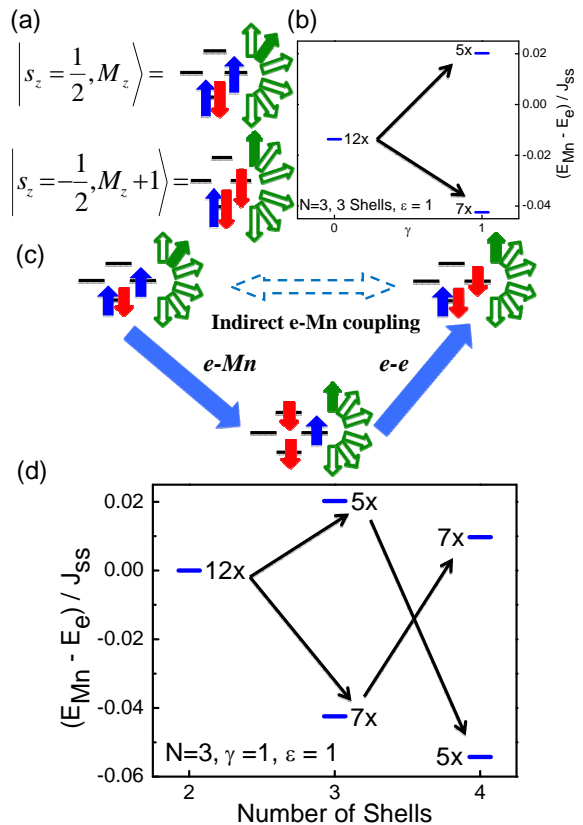}
\caption{(Color online) (a) Ground-state three-electron
  configurations with the $p$-shell electron spin up (top) and spin
  down (bottom).
  (b) Energy difference $\Delta$ between a three-electron GS in the
  Mn-doped and undoped QD for both noninteracting ($\gamma=0$) and
  interacting ($\gamma=1$) electrons. 
  The numbers indicate the degeneracy of each level.
  (c) Diagram of coupling between electrons in the $p$ shell and Mn. 
  The solid arrow represents a direct coupling via $e$-Mn coupling or
  $e$-$e$ Coulomb interaction, and the dashed arrow illustrates the
  indirect coupling.
  (d) The energy difference $\Delta$ as a function of the number of shells
  for a QD containing three shells and $\gamma=1$. \label{fig2}}
\end{center}
\end{figure}

Figure~\ref{fig2}(a) illustrates the degenerate three-electron
configurations, $|s_{z}=1/2,M_{z} \rangle$ and $|s_{z}=-1/2,M_{z}+1
\rangle$, with an electron with spin up and Mn in state $M_z$ and
and electron with spin down on the $p$ orbital and Mn in state $M_z+1$. 
As the electron-Mn exchange interaction in the $p$ shell vanishes,
$J_{pp}=0$, these configurations do not interact with each other. 
As a consequence, the GS is 12-fold degenerate, two electron spin
configurations times six Mn spin orientations. 
In order to understand the effect of interactions we include
configurations coupled with $|s_{z}=1/2,M_{z} \rangle$ and 
$|s_{z}=-1/2,M_{z}+1\rangle$ by both $e$-$e$ and $e$-Mn interactions and
diagonalize the Hamiltonian in the $L=1$ subspace. 
The number of three-electron-Mn configurations depends on the number
of electronic shells, with  $24$, $228$, $852$, and $2520$ for two, three,
four, and five shells, respectively.  
%%%%%%%%%%%%%%%%%%%%%%%%%%%%%%%%%Figure 2 N=3 configurations %%%%%%%%
%%%%%%%%%%%%%%%%%%%%%%%%%%%%%%%%%%%%%%%%%%%%%%%%%%%%%%%%%%%%%%

Figure \ref{fig2}(b) shows the result of exact diagonalization of the
$e$-Mn Hamiltonian for three confined shells in the QD and an isotropic
$e$-Mn interaction ($\epsilon=1$), for both noninteracting  
($\gamma=0$) and interacting ($\gamma=1$) electron systems. 
For the noninteracting case we observe that the GS is 12-fold 
degenerate, with the energy lowered by the $e$-Mn interaction (negative
$\Delta$).
This behavior is identical to what was shown for the two electrons in
the previous section, i.e., the two electrons in the $s$ shell are coupled
with Mn, while the electron in the $p$ shell is only a spectator. 
However, in the strongly interacting regime, $\gamma=1$, we observe a
splitting of the degenerate GS into two degenerate shells. 
The splitting and the degeneracy of  levels is consistent with an
effective Hamiltonian $H_{\text{eff}}=-J_{\text{eff}} \vec{s} \cdot \vec{M}$ coupling
the $p$-shell electron spin $s$ with Mn spin
$M$.\cite{qu_hawrylak_prl2005} 
The effective coupling $J_{\text{eff}}$ is mediated by Coulomb interactions.
In Fig.~\ref{fig2}(c) we illustrate the processes which couple
$|s_{z}=1/2,M_{z} \rangle$ and $|s_{z}=-1/2,M_{z}+1\rangle$ states. 
The $e$-Mn interaction acting on the $|s_{z}=1/2,M_{z} \rangle$ state
scatters the spi- up (blue) electron from the $s$ shell to the spin-down 
(red) electron on the $d$ shell with a simultaneous transition of
the Mn spin from $M_z$ to $M_z+1$. 
In the next step, the $e$-$e$ interaction scatters the $d$-shell and
$p$-shell electron pair into the $s$-shell and $p$-shell electron
pair, with the spin-down electron on the $p$ shell and the spin-up
electron on the $s$ shell. 
The net result is a spin flip of the $p$-shell electron and of
the Mn spin. 
We see that the ground state is sevenfold degenerate, implying that the
electron spin is aligned with the Mn spin and $J_{\text{eff}}$ is hence
ferromagnetic.  

Let us now investigate the dependence of the GS energy on the number
of confined shells in the QD. 
Figure~\ref{fig2}(d) shows the evolution of the GS energy as a function
of the number of shells for $\gamma=1$ and $\varepsilon=1$. 
We observe that for two shells there is no splitting ,
i.e., $J_{\text{eff}}=0$, while for three and four shells the GS is split
into two shells.  
For two shells the GS is 12-fold degenerate, $\Delta=0$, and there
is no interaction between Mn and electrons.  
For three shells the GS is split into two shells, as discussed above.
For four shells the GS is also split into two, but there is 
an inversion of the  degeneracy of the energy levels. 
This is a consequence of an antiferromangetic interaction $J_{\text{eff}}<0$
between the electron and Mn spins. 
We also have observed that for QDs confining five or six shells the
results are similar to what was obtained for QD with four shells, i.e.,
the antiferromagetic coupling is stabilized for a QD containing more
than three confined shells. 
This can be understood by looking at the way the GS is coupled to Mn. 
In Fig.~\ref{fig2}(c) we show that there is an indirect coupling
between configurations $|s_{z}=1/2,M_{z}\rangle$ and
$|s_{z}=-1/2,M_{z}+1 \rangle$ which is mediated via $e$-$e$ Coulomb and
$e$-Mn interactions between the GS and excited configurations.  
As the number of shells increases, more excited state configurations
interact with the GS, stabilizing the antiferromagnetic indirect
coupling between the electrons and Mn.  

\begin{figure}
\begin{center}
\includegraphics[width=0.48\textwidth]{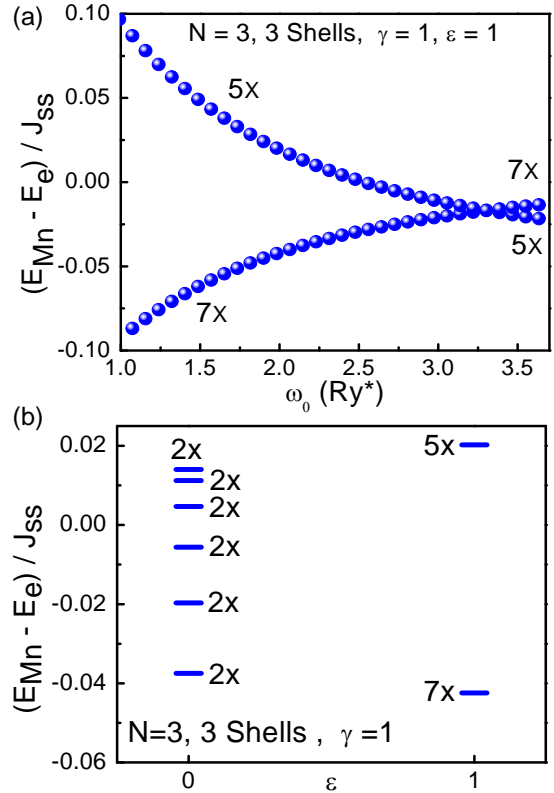}
\caption{(Color online) 
  (a) Evolution of the energies of three-electron levels with $J=3$
  and $J=2$ as a function of the QD shell spacing $\omega_{0}$.  
  (b) Energy difference $\Delta$ for both anisotropic
  ($\varepsilon=0$) and isotropic ($\varepsilon=1$) $e$-Mn interactions
  in a three-shell QD with full interactions ($\gamma=1$). 
\label{fig3}}
\end{center}
\end{figure}

If the indirect magnetic ordering shown above depends on the number of
shells, it also should depend on the QD shell spacing $\omega_{0}$.  
Figure~\ref{fig3}(a) shows the dependence of GS energy on $\omega_{0}$
for three electrons confined in a Mn-doped QD containing three shells, 
$\gamma=1$ and $\varepsilon=1$. We note that the exchange coupling 
changes from ferromagnetic to antiferromagnetic  for $\omega_{0} \approx 3.3$ 
Ry$^{*}$. We observe the same behavior for QDs with four shells, 
but in this case the crossing occurs at $\omega_{0} \approx 0.45$ Ry$^{*}$.  
%%%%%%%%%%%%%%%%%%%%%%%%%%%%%%%%% Figure 3 %%%%%%%%%%%%%%%%%%%
%%%%%%%%%%%%%%%%%%%%%%%%%%%%%%%%%%%%%%%%%%%%%%%%%%%%%%%%%%%%%%%%

Next we discuss the effect of anisotropy on the $e$-Mn exchange
interaction. Figure~\ref{fig3}(b) shows the GS energy for three 
electrons in a QD containing three shells in the strongly interacting 
regime as a function of the $e$-Mn coupling. For $\epsilon=0$ the 
electrons and Mn interact via an anisotropic Ising-like Hamiltonian, 
and for $\epsilon=1$ the $e$-Mn interaction is isotropic, Heisenberg-like. 
For $\epsilon=0$, $s_{z}$ is a good quantum number, and therefore the 
electron spin degeneracy is preserved. In Fig.~\ref{fig3}(b) we observe 
that for $\epsilon=0$ the energy spectrum is split into six  doubly 
degenerate levels. This splitting is due to the $e$-$e$ Coulomb interaction 
driving the indirect $e$-Mn interaction between the $p$-shell electron and 
Mn, as was observed in the $\epsilon=1$ case. The double degeneracy 
for the anisotropic coupling arises due to the fact that the state 
$|s_{z}=1/2,M_{z} \rangle$ has the same energy as the configuration of 
$|s_{z}=-1/2,-M_{z} \rangle$. 

We also investigated the effect of Mn positions on three-electron GSs. 
Moving Mn away from the QD center couples the electron in the $p$ orbital 
directly with Mn. This coupling is ferromagnetic. Considering a QD containing 
three shells and $\omega = 1.98$ Ry$^{*}$, the indirect $e$-Mn coupling 
is also ferromagnetic, and therefore both direct and indirect $e$-Mn 
interactions add up. As Mn is moved away from the QD center, the direct 
coupling becomes the dominant effect for Mn positions larger than 
$R \approx 0.2 l_{0}$. Even though the direct $e$-Mn interaction is 
dominant for Mn far away of the QD center, the indirect $e$-Mn coupling 
is always present.
 
%%%%%%%%%%%%%%%%%%%%%%%%%%%%%%%%%%%%%%%%%%%%%%%%%%%%%%
\subsection{Two spin-polarized electrons on the $p$ shell}

\begin{figure}
\begin{center}
\includegraphics[width=0.48\textwidth]{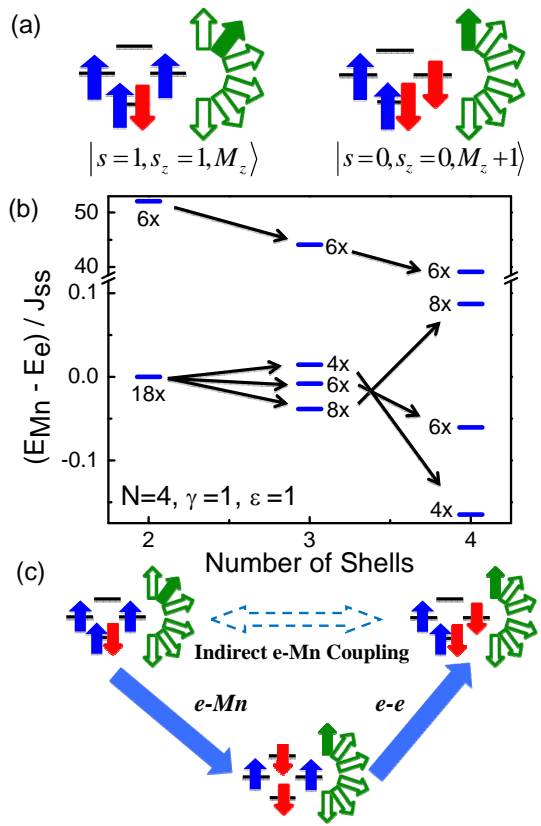}
\caption{(Color online) 
  (a) Low-energy configurations of four electrons in a magnetic QD.
  (b) Low-energy spectrum of the system as a function of the number 
  of shells for interacting ($\gamma=1$) electrons, measured from
  the respective GS energy $E_e$ of a nonmagnetic system. 
  Here the QD shell spacing $\omega_{0}=1.98$ Ry$^{*}$. 
 (c) Indirect coupling diagram of two four-electron configurations.
 The solid arrows represent direct interactions between configurations
 and the dashed arrow represents the indirect $e$-Mn coupling. \label{fig4}}
\end{center}
\end{figure}

Next we describe the electronic properties of a half-filled $p$ shell.
The lowest-energy configuration of the four-electron GS state is 
formed by two electrons in the $s$ shell and two spin triplet
electrons in the $p$ shell. Figure~\ref{fig4}(a) illustrates the 
four-electron configurations, the triplet $|S=1,s_{z}=1,M_{z}\rangle$ and 
one of the singlet component $|S=0,s_{z}=0,M_{z}+1\rangle$ configurations.  
These two configurations have the same total spin projection $J_z$.
In the presence of an $e$-$e$ Coulomb interaction the $S=1$ triplet state is the GS 
and the singlet is an excited state. 
For Mn in the QD center the $p$ electrons do not couple with Mn,  the
electron spin degeneracy is preserved, and the degeneracy of the
triplet state in a Mn-doped QD is 18, while the singlet state is
sixfold degenerate.  

We shall now investigate how the GS of four electrons confined in a
Mn-doped QD is affected by the presence of the $e$-$e$ Coulomb interaction,
number of shells, shell spacing, and $e$-Mn coupling. 
We take advantage of the conservation of the total angular momentum
and diagonalize our microscopic Hamiltonian in the $L=0$ subspace. 
The number of configurations in this subspace is $30$, $498$, and
$3498$ for two, three, and four shells, respectively.  
%%%%%%%%%%%%%%%%%%%%%%%%%%%%%% Figure 4 %%%%%%%%%%%%%%%%%%
%%%%%%%%%%%%%%%%%%%%%%%%%%%%%%%%%%%%%%%%%%%%%%%%%%%%%%

The $e$-$e$ mediated coupling of the electronic and Mn spin is interpreted in
terms of the effective exchange Hamiltonian. 
Adding the electron and Mn spins results in total spin $J=7/2,5/2,3/2$
and splitting of the 18-fold degenerate ground state into eightfold, sixfold
and fourfold degenerate shells. 
Figure~\ref{fig4}(b) shows the evolution of the low-energy part of the
spectrum of four electrons in the magnetic dot as a function of the
number of shells for full $e$-$e$ interactions ($\gamma=1$) and the
isotropic $e$-Mn coupling ($\varepsilon=1$).
The energies of these states are shown relative to the energy of the
ground-state triplet of the undoped QD.
The triplet and singlet states split for any number of shells due to
an $e$-$e$ exchange interaction. 
In a QD with only $s$ and $p$ shells, the effective exchange coupling
for $p$-shell electrons is zero and the triplet and singlet states are
18 and six times degenerate, respectively. 
Increasing the number of shells leads to a finite and ferromagnetic
exchange interaction with the triplet states coupled to the Mn spin and
the 18-fold degenerate shell split into eight-, six- and four-fold degenerate
levels.  
The character of this exchange interaction depends on the number of
shells. 
For three shells we have a ferromagnetic coupling, but for four shells
the coupling becomes antiferromagnetic. 

Figure~\ref{fig4}(c) illustrates the configurations involved in the
indirect coupling of the electrons on the $p$ shell and the Mn spin.
Here, the solid arrows represent the direct coupling between
configurations, and the dashed arrow represents the indirect
interaction between two configurations. 
Let us explain how this indirect coupling arises, starting from the
configuration with two spin-up electrons in the $p$ shell, which is
labeled as $|S=1,s_{z}=1,M_{z} \rangle$ [see Fig.~\ref{fig4}(c), top left]. 
This configuration is coupled with an excited state in which there are
two spin-down electrons in both $L_{e}=0$ orbitals, one in  
the $s$ shell and the other in the $d$ shell. 
This coupling occurs via an $e$-Mn interaction, which scatters the spin-up
electron in the $s$ shell of $|S=1,s_{z}=1,M_{z} \rangle$ to the
$d$ shell, flipping the electron spin down, and the Mn spin up, i.e.,
$M_{z}+1$.
This  excited state with $s_{z}=0$ and $M_{z}+1$ is coupled with one
of the $|S=0,s_{z}=0,M_{z}+1 \rangle$ GS configurations via  the $e$-$e$
Coulomb interaction, in which the spin-down electron in the
$d$ shell is scattered to the $L_{e}=1$ $p$ orbital, and the spin-up  
electrons in this orbital are scattered to the $s$ shell. 

\begin{figure}
\begin{center}
\includegraphics[width=0.55\textwidth]{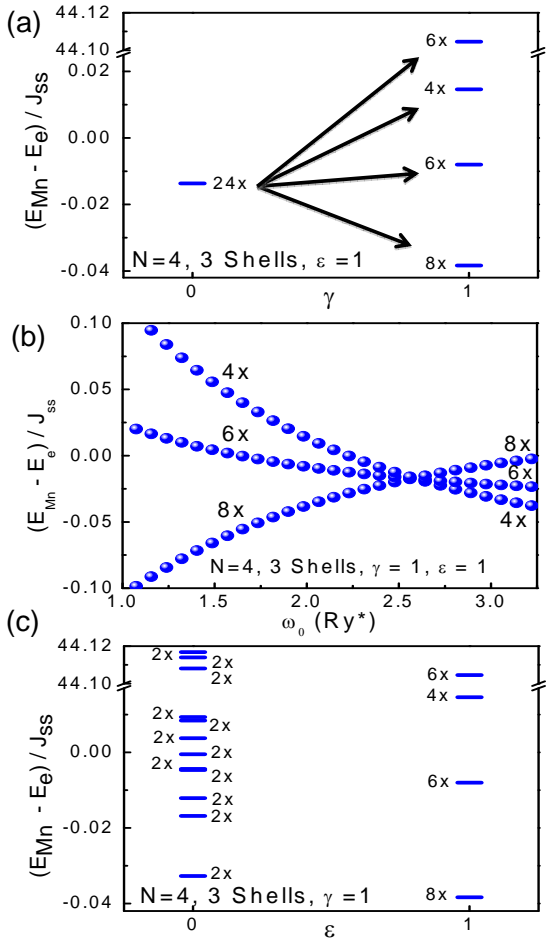}
\caption{(Color online) 
  (a) Energy difference $\Delta$ for noninteracting ($\gamma=0$) and
  interacting ($\gamma=1$) electrons in the four-electron magnetic dot. 
  (b) GS energy difference as a function of the QD shell spacing
  $\omega_{0}$ for three shells confined in the QD. 
  (c) GS energy difference for the anisotropic ($\varepsilon=0$) and
  isotropic ($\varepsilon=1$) $e$-Mn coupling. \label{fig5}} 
\end{center}
\end{figure}

Figure~\ref{fig5}(a) shows the GS energy for both noninteracting
($\gamma=0$) and fully interacting ($\gamma=1$) electrons. 
We considered a Mn-doped QD with three confined shells and the
isotropic $e$-Mn interaction ($\varepsilon=1$). 
For the noninteracting case there is no triplet-singlet splitting, and
as and $e$-$e$ Coulomb interaction mediates the indirect interaction between
Mn and the $p$-shell electrons, the triplet is not split either.
Therefore, the four-electron GS is 24-fold degenerate.
Even though the four noninteracting electron triplet states are not
split by the indirect coupling, we see a negative $\Delta$, which means
that electrons  lower their energy by an exchange interaction with Mn. 
Turning the $e$-$e$ Coulomb interaction on results in the singlet-triplet
splitting and a further splitting of the triplet energy shell.
The triplet splitting is caused by the indirect interaction between Mn
and electrons in the $p$ shell, which is mediated by the $e$-$e$ Coulomb
interaction. 
%%%%%%%%%%%%%%%%%%%%%%%%%%%%%% Figure 5 %%%%%%%%%%%%%%%
%%%%%%%%%%%%%%%%%%%%%%%%%%%%%%%%%%%%%%%%%%%%%%%%%%%%%%

In Fig.~\ref{fig5}(a) we  show the effect of the $e$-$e$ Coulomb
interaction on the low-energy spectrum of the four-electron and Mn
complex.
We note the appearance of  triplet and singlet energy shells, separated
by the $e$-$e$ exchange interaction. 
The splitting of the triplet shell is governed by the $e$-$e$ and $e$-Mn
exchange interactions.  

Figure~\ref{fig5}(b) presents the energy difference $\Delta$, i.e., the
effective exchange coupling, as a function of $\omega_{0}$ for four
interacting electrons ($\gamma=1$) confined in the Mn-doped QD with
three confined shells.  
Here we also have a ferromagnetic to antiferromagnetic crossing as a
function of the QD shell spacing. 
For QDs with four shells the  ferromagnetic to antiferromagnetic
crossing occurs at $\omega_{0} \approx 0.04$ Ry$^{*}$.  

Now we show the effect of the symmetry of the $e$-Mn coupling on the
four-electron GS. 
In Fig.~\ref{fig5}(c) we compare the effects of the anisotropic 
($\varepsilon=0$) and isotropic ($\varepsilon=1$) coupling for a
Mn-doped QD with three confined shells and in the presence of a full
$e$-$e$ Coulomb interaction ($\gamma=1$). 
For the anisotropic coupling the triplet state is split into nine
doubly degenerate levels. 
In this case, both $s_{z}$ and $M_{z}$ are good quantum numbers, and
therefore, $s_{z}=1$ and $s_{z}=-1$ breaks the Mn spin degeneracy into
six. 
As the energy of the state with $s_{z}=1$ and $M_{z}$ is equal to the
energy of the state $s_{z}=-1$ and $-M_{z}$, these six states are
double degenerate. 
The $s_{z}=0$ configurations split into three, where the degeneracy 
is given by $M_{z}$, i.e., the $s_{z}=0$ configurations are degenerate
and labeled by $|M_{z}|$, as for the two electrons interacting with the
Mn via an anisotropic $e$-Mn interaction. 
The singlet state  is also split into three doubly degenerate levels.  

One way to probe the indirect $e$-Mn interaction is by 
performing a circularly polarized photoluminescence experiment of quantum 
dots containing a single Mn spin and confining three or more electrons. 
In this case, the indirect $e$-Mn coupling gives rise to a fine structure 
of both initial and final states of the emission process.
\cite{mendes_korkusinski_prb2013}

\section{Conclusion}

In conclusion, we presented  a microscopic model of interacting
electrons coupled with a magnetic ion spin localized 
in the center of a self-assembled quantum dot.  
We showed that the electrons occupying finite angular momentum orbitals
interact with the localized spin through an effective exchange
interaction mediated by electron-electron interactions.  
The effective interaction for $p$-shell electrons is obtained using
exact diagonalization of the microscopic Hamiltonian as a function of 
the number of electronic shells, shell spacing, and anisotropy of the 
exchange interaction. 
It is shown that the effective interaction can be engineered to be
either ferro- or antiferromagnetic, depending on the quantum-dot
parameters. 

\section*{ACKNOWLEDGMENT}

The authors thank NSERC and the Canadian Institute for Advanced
Research for support. UCM thanks J. A. Brum for fruitful discussions and 
acknowledges the support from CAPES-Brazil (Project No. 5860/11-3) 
and FAPESP-Brazil (Project No. 2010/11393-5). PH thanks Y. Hirayama, 
WPI-AIMR, Tohoku University for hospitality.

%\bibliographystyle{apsrev4-1}
%\bibliography{References_QDPawel}

\end{document}